\documentclass[a4paper]{article}

\usepackage{INTERSPEECH2021}
\usepackage{booktabs}
\usepackage{multirow}
\usepackage{cite}
\usepackage{placeins}

\title{NISQA: A Deep CNN-Self-Attention Model for Multidimensional Speech Quality Prediction with Crowdsourced Datasets}
\name{Gabriel Mittag$^1$, Babak Naderi$^1$, Assmaa Chehadi$^1$, Sebastian M\"oller$^{1,2}$}
\address{$^1$Quality and Usability Lab, Technische Universit\"at Berlin, Berlin, Germany \\
$^2$Deutsches Forschungszentrum f\"ur K\"unstliche Intelligenz (DFKI), Berlin, Germany }
\email{first.last@tu-berlin.de}

\begin{document}

\maketitle
\begin{abstract}
In this paper, we present an update to the NISQA speech quality prediction model that is focused on distortions that occur in communication networks. In contrast to the previous version, the model is trained end-to-end and the time-dependency modelling and time-pooling is achieved through a Self-Attention mechanism. Besides overall speech quality, the model also predicts the four speech quality dimensions \textit{Noisiness}, \textit{Coloration}, \textit{Discontinuity}, and \textit{Loudness}, and in this way gives more insight into the cause of a quality degradation. Furthermore, new datasets with over 13,000 speech files were created for training and validation of the model. The model was finally tested on a new, live-talking test dataset that contains recordings of real telephone calls. Overall, NISQA was trained and evaluated on 81 datasets from different sources and showed to provide reliable predictions also for unknown speech samples. The code, model weights, and datasets are open-sourced.
\end{abstract}

\section{Introduction}
One of the main performance indicators for the evaluation of telecommunication networks is the perceived speech quality. It is traditionally derived from subjective listening tests according to ITU-T P.800 \cite{P800} or recently also through crowdsourced listening tests according to ITU-T P.808 \cite{P808, p808imp}. The average rating across all test participants for a speech sample then gives the mean opinion score (MOS). However, because listening tests are costly and time consuming, instrumental models have been developed that can predict the speech quality automatically. The currently recommended model by the ITU-T is POLQA \cite{P863}, which requires the clean reference and the degraded output signal to predict the speech quality based on a comparison of both signals. In contrast to these types of \textit{double-ended} models, \textit{single-ended} models only require the degraded output signal, which makes it possible to monitor the quality of live phone calls, or predict the speech quality of samples for which no clean reference is available. However, the currently recommended single-ended speech quality by the ITU-T, P.563 \cite{p563}, is only available for narrowband (NB) speech signals. Furthermore, the prediction performance showed to decrease for conversational speech and modern VoIP (Voice over IP) distortions that were not present when P.563 was developed \cite{Hines2015MeasuringAM}. 

While the overall MOS is an important indicator, it gives no insight into the cause of a quality degradation. To overcome this problem, it was shown in \cite{wal} that the speech quality multidimensional space of modern communication networks is made up of the three orthogonal dimensions: \textit{Noisiness}, \textit{Coloration}, and \textit{Discontinuity}. Later, the \textit{Loudness} was added as a fourth dimension in \cite{cote_loudness}, although it is not entirely orthogonal to the other dimensions. These four perceptual dimensions can be quantified through auditory listening tests and are linked to technical root causes.

Recently, deep learning methods have been applied to build single-ended speech quality models \cite{Soni2016, Ooster2019, Fu2018, wenet, Dong_1, Dong_2, Avila2019, Cauchi, Serr2020SESQASL} and showed to outperform traditional approaches without the need for a clean reference. In \cite{mittag2019ic}, we presented the deep learning model NISQA that predicts speech quality of super-wideband (SWB) speech samples. In \cite{mittag2019dim}, we presented an extension of the model that predicts three of the four quality dimensions based on expert scores due to the lack of available subjective data. In this paper, we present an update to NISQA that predicts the overall MOS and the four speech quality dimensions with one multi-task neural network. Furthermore, we created a large pool of eight new speech quality datasets with subjective MOS and quality dimension ratings for training and evaluation. Because of the increased available data, we could train the model end-to-end with subjective data only, without the need for objective MOS values. Also, we improved the neural network architecture of the model by replacing the CNN-LSTM structure with a CNN--Self-Attention--Attention-Pooling (CNN-SA-AP) network. The model is overall trained and evaluated on a large set of 81 datasets from different sources. Another advantage of the updated NISQA model is that it can be applied to speech samples of any duration or sample rate without any preprocessing steps or level normalisation. Finally, the PyTorch code, the model weights, and several of the datasets are open-sourced on GitHub\footnote{www.github.com/gabrielmittag/NISQA}.

\section{Method}
\begin{figure}[ht]
    \centering
    \centering\includegraphics[width=0.75\linewidth]{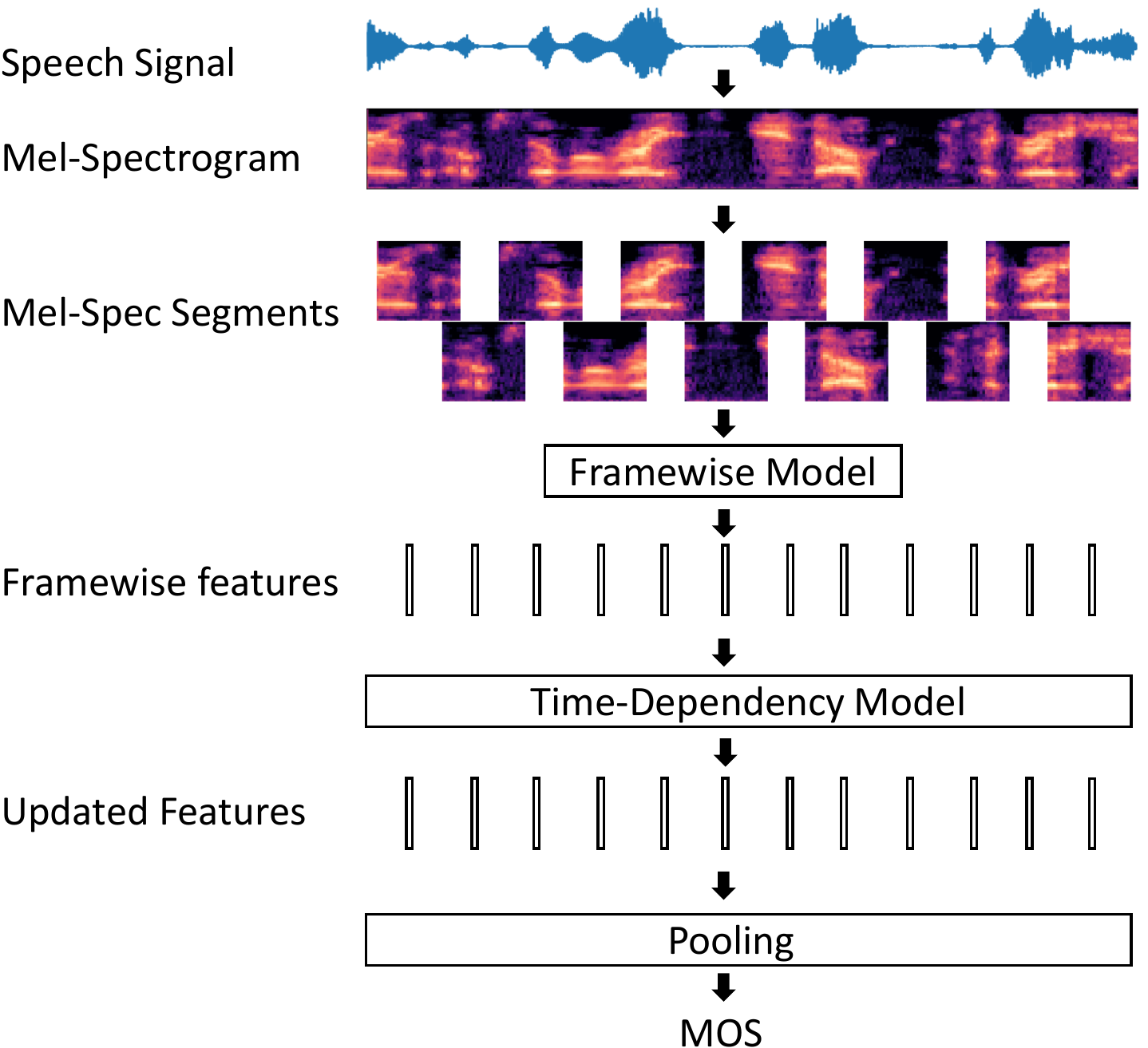} 
\caption{General speech quality model structure.}
\label{fig:nn_model}
\end{figure}

The model can be divided into four stages: 1) Mel-Spec segmentation, 2) Framewise model (CNN), 3) Time-Dependency model (Self-Attention), 4) Pooling model (Attention-Pooling). An overview of this architecture is shown in Figure \ref{fig:nn_model}. In the first stage, Mel-specs are calculated from the input signal and then divided into overlapping segments. In the second stage, a framewise neural network with the Mel-spec segments as inputs is used to compute features that are suitable for speech quality prediction. These features are calculated on a frame basis and therefore result in a sequence of framewise features. In a third stage, the time dependencies of the feature sequence are modelled. Finally, the features are aggregated over time in a pooling layer. The aggregated features are then used to estimate a single MOS value. 

In this paper, different neural network architectures are applied and compared for each model stage. In the ablation study (Sect.~\ref{sec:abl}), we show that the best combination is a CNN as a framewise model, a Self-Attention network as a time-dependency model, and Attention-Pooling as a pooling model. More details about the neural network structure can be found in the open-sourced code.

\subsection{Mel-Spec Segmentation}
The input to the model are Mel-spec segments with 48 Mel-bands. The FFT window length is 20\,ms with a hop size of 10\,ms. The maximum frequency was chosen to be 20\,kHz to be able to predict speech quality for up to fullband (FB). The Mel-specs are divided into segments with a width of 15 (i.e. 150\,ms) and a height of 48. The hop size between the segments is 4 (40\,ms), which leads to a segment overlap of 73\% and overall 250 segments for a 10-second speech signal. The framewise network is provided with this wider segment of 150\,ms to give the network some contextual awareness. The short-term and long-term temporal modelling, however, follows in the third model stage.

\subsection{Framewise Model}
As framewise model the CNN from \cite{mittag2019ic} is used. It contains 6 convolutional layers and 3 max-pooling layers that downsample the input of dimension $48\times15$ to a size of $6\times3$. The final convolutional layer does not apply a width padding and therefore further reduces the output dimension to $64\times6\times1$, where 64 represent the number of kernels. Finally, the output is flattened. Thus, each Mel-spec segment results in a feature vector of length 384 after passing the CNN. As a baseline model, a basic deep feedforward network with a depth of four layers and 2048 hidden units each is implemented. 
\subsection{Time Dependency}
 In this stage, the individual time steps of the feature sequence can interact with each other to improve the prediction performance. To this end, a Self-Attention network is applied, which is based on the Transformer encoder \cite{NIPS2017_7181}. Because the Self-Attention only models temporal dependencies of framewise features that are already computed by a deep framewise network, a relatively low complexity of the Transformer is sufficient. Moreover, in practice, it was noted that the multi-head mechanism did not improve the results. Therefore, the block is implemented with a single head, a depth of 2 blocks, a model dimensionality of $d_\mathrm{tf}=64$, and a feedforward network with $d_\mathrm{tf,ff}=64$ hidden units. As a baseline, a single BiLSTM layer with 128 hidden units in each direction is used.

\subsection{Pooling}
\begin{figure}[ht]
    \centering
    \centering\includegraphics[height=4cm]{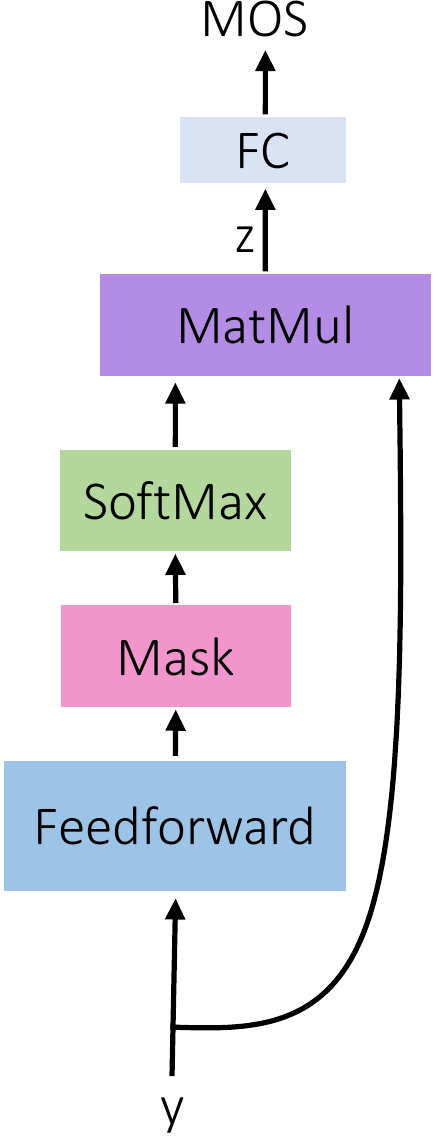} 
\caption{Attention-pooling block}
\label{fig:att_pool}
\end{figure}

The quality of a telephone call can generally not be predicted accurately by simply taking the average quality across time. As was shown in \cite{berger2008a}, the ``recency effect'' and the out-weighting of poor quality segments in a call have to be considered adequately. Therefore, we propose to use an attention mechanism for time pooling. An overview of the attention-pooling block is shown in Figure~\ref{fig:att_pool}, where $y$ is the output $d_\mathrm{tf}$x$L$-matrix of the time-dependency model. The matrix contains a zero-padded sequence of length $L$ with feature vectors of dimension $d_\mathrm{tf}$. The feedforward network with an output size of 1 and 128 hidden units is applied to each time step separately and identically. The attention scores computed by the feedforward network are then masked at the zero-padded time steps and applied to a softmax function to yield the normalised attention weights. These weights are applied to the input matrix $y$ with a matrix multiplication operation. The weighted average feature vector $z$ is then finally passed through a fully connected layer to estimate the overall speech quality. As baseline models, \textit{average-pooling} and \textit{max-pooling} are applied.

\subsection{Multidimensional model}
\begin{figure}[ht]
    \centering
    \centering\includegraphics[width=0.7\linewidth]{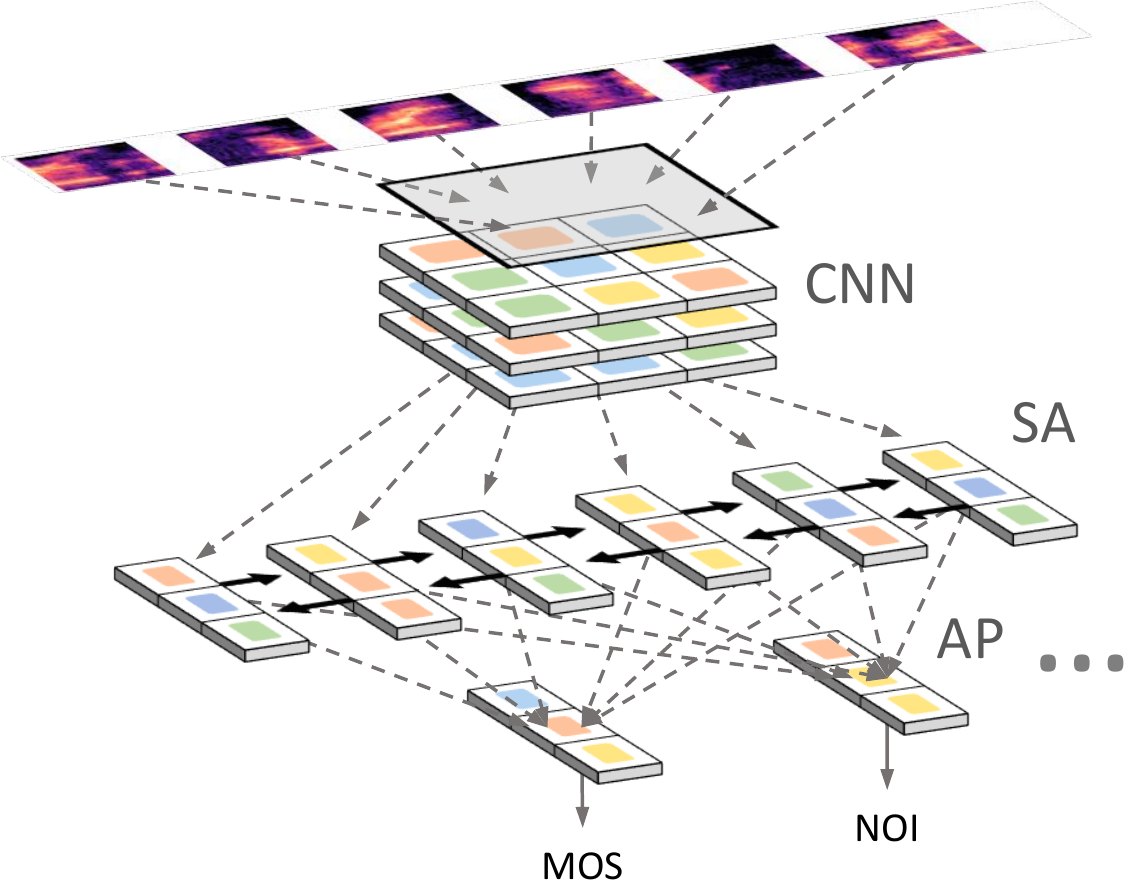} 
\caption{NISQA neural network architecture.}
\label{fig:mt_model_pool}
\end{figure}

The multidimensional prediction can be seen as a Multi-Task-Learning problem. The pooling block is computed separately for each dimension and overall MOS, while the CNN and Self-Attention network is shared across all tasks. Figure \ref{fig:mt_model_pool} shows how the Mel-spec features are calculated by the same CNN and Self-Attention network for each dimension. The outputs of each Self-Attention time-step are then the input for five individual pooling blocks that predict the overall MOS and the dimension scores. 

\section{Datasets}
NISQA is trained and evaluated on a large set of 59 training datasets (72,903 files), 18 validation sets (9,567 files), and 4 test sets (952 files)\footnote{A detailed overview of the individual datasets can be found on the GitHub.}. 55 of the datasets are taken from the POLQA Pool \cite{P863}, 7 of the datasets are taken from the ITU-T P Suppl. 23 \cite{supp23} pool, 11 datasets are older, internal speech quality datasets. Additionally, for this work 8 new datasets with overall quality and quality dimension ratings, and a large variety of different speakers, were created. 2 training datasets NISQA\_TRAIN\_SIM (10,000 samples from 2,322 speakers), NISQA\_TRAIN\_LIVE (1,020 samples from 486 speakers) and 2 validation datasets NISQA\_VAL\_SIM (2,500 samples from 938 speakers), NISQA\_VAL\_LIVE (200 samples from 102 speakers). The clean source speech samples are taken from four different English speech corpora: AusTalk \cite{austalk} (containing conversational speech taken from interviews), DNS-Challenge \cite{reddy2020interspeech} (LibriVox audiobooks), TSP \cite{tspdb} (read out Havard sentences \cite{harvard}), and UK-Ireland dataset \cite{google_dataset} (read out public domain texts). The datasets NISQA\_TRAIN\_SIM and NISQA\_VAL\_SIM contain simulated speech distortions, such as packet-loss, bandpass filter, different codecs, and clipping. To simulate real background noises, the noise clips from the DNS-Challenge datasets were used, which in turn are taken from the Audioset \cite{audioset}, freesound \cite{freesound}, and DEMAND \cite{demand} corpora. The datasets NISQA\_TRAIN\_LIVE and NISQA\_VAL\_LIVE contain live Skype and landline-to-mobile phone recordings, where the LibriVox audiobook reference files were played back through a loudspeaker directly into the terminal device (phone/laptop). During the recording different real distortions were created, such as typing on a keyboard, opening window (street noise), or poor reception. These four training and validation datasets were annotated in the crowd according to ITU-T P.808 \cite{P808} with 5 ratings per file.

Additionally, four independent test sets were created that were not considered before the final training of the NISQA model. NISQA\_TEST\_P501, NISQA\_TEST\_FOR, NISQA\_TEST\_NSC contain simulated distortions and additionally live VoIP calls with Zoom, Skype, Google Meet, WhatsApp, and Discord, where the reference speech samples were played back directly from the laptop. Then a poor internet connection was simulated to obtain files with different distortions (packet-loss, warping, low-bitrate). NISQA\_TEST\_P501 contains the English Annex C files from ITU-T P.501 \cite{P501}, NISQA\_TEST\_FOR contains English conversation samples from the Forensic Voice dataset \cite{Morrison2012ProtocolFT, for_database}, NISQA\_TEST\_NSC contains German conversational speech from the NSC dataset \cite{Gallardo}. These three datasets were again annotated in the crowd according to ITU-T P.808 with 30 ratings per file.

Because the use-case for a single-ended prediction model are real phone calls with conversational speech, from an unknown speaker, device, and network, a fourth dataset NISQA\_TEST\_LIVETALK with real phone call recordings was created, where the talkers spoke directly into the terminal device (i.e. a smartphone or laptop). The test participants were instructed to talk loudly, quietly, with loudspeaker, or music in the background to obtain different test scenarios and speech quality distortions. Depending on the condition the talkers were located in different environments, such as in a café, inside a car on the highway, inside a building with poor reception, elevator, shopping centre, subway/metro station, on a busy street, etc. The talkers used their mobile phone to call either through the mobile network or with a VoIP service (Skype/Facebook). The dataset consists of 58 different conditions with each 4 different files, resulting in 232 files overall. The speech files were recorded from 8 different talkers (4 male and 4 female) in German, where for each condition 2 male and 2 female talkers were selected. The dataset was annotated in the lab according to P.800 with 24 ratings per file.

\section{Ablation Study}
\label{sec:abl}
In this section, it will be shown that a combination of a CNN as a framewise model, a Self-Attention network as time-dependency model, and an Attention-Pooling network as pooling model (CNN-SA-AP) gives the highest prediction performance. To this end, an ablation study is performed, in which one of the three neural network model stages is either removed or replaced with another network type. For training, the simulated and live training sets NISQA\_TRAIN\_SIM and NISQA\_TRAIN\_LIVE were used. The models are evaluated on the average PCC (Pearson's correlation coefficient) between the validation datasets NISQA\_VAL\_SIM and NISQA\_VAL\_LIVE. The training for each model configuration is run 12 times to rule out random effects, the median performance over these 12 training runs is presented in the following result tables.

\subsection{Framewise Model}
Table~\ref{tab:framewise} shows the results when different framewise models are applied. The \textit{CNN}-SA-AP model clearly outperforms the basic feedforward neural network (\textit{FFN}-SA-AP) and the model without framewise network that only applies Self-Attention and Attention-Pooling (\textit{Skip}-SA-AP). 

\begin{table}[!htb]
\footnotesize
\caption{Framewise model comparison with Median PCC.}
\centering
\label{tab:framewise}
\begin{tabular}{@{}lccc@{}}
\toprule
Model      & Skip  & CNN           & FFN    \\ \midrule
\textit{r} & 0.772 & \textbf{0.870} & 0.812  \\ \bottomrule
\end{tabular}
\end{table}

\subsection{Time-Dependency Model}
Table~\ref{tab:td} shows the results for different time-dependency models. Again, it can be seen that the CNN-\textit{SA}-AP model achieves the best results. However, the difference between Self-Attention and CNN-\textit{LSTM}-AP is only small. Further, it can be seen that a combination of SA and LSTM worsens the results, compared to SA or LSTM only. Although the difference between SA and LSTM is small, the overall performance can be notably increased when compared to the model CNN-\textit{Skip}-AP without time-dependency modelling.

\begin{table}[htb]
\footnotesize
\caption{Time-Dependency comparison with Median PCC.}
\centering
\label{tab:td}
\begin{tabular}{lccccc}
\hline
Model      & Skip  & SA            & LSTM  & LSTM-SA & SA-LSTM \\ \hline
\textit{r} & 0.851 & \textbf{0.870} & 0.866 & 0.862   & 0.863   \\ \hline
\end{tabular}
\end{table}

\subsection{Pooling}

\begin{table}[htb]
\footnotesize
\caption{Pooling comparison with Median PCC.}
\centering
\label{tab:pool}
\begin{tabular}{lccc}
\hline
Model      & AP   & Avg            & Max   \\ \hline
\textit{r} & \textbf{0.870} & 0.867 & 0.866 \\ \hline
\end{tabular}
\end{table}

\begin{table*}[!htbp]
\caption{Per-condition validation and test results of the overall quality in terms of PCC and RMSE after first-order mapping.}
\centering
\label{tab:results_val_de}
\resizebox{\textwidth}{!}{%
\begin{tabular}{@{}lp{0.5cm}p{0.4cm}rrccccccccccccccccccccc@{}}
\toprule
Dataset              & Scale & Lang & Con  & Files &  & \multicolumn{2}{c}{NISQA} &  & \multicolumn{2}{c}{P563} &  & \multicolumn{2}{c}{ANIQUE+} &  & \multicolumn{2}{c}{WAWEnets} &  & \multicolumn{2}{c}{POLQA} &  & \multicolumn{2}{c}{DIAL} &  & \multicolumn{2}{c}{VISQOL}\\
 \cmidrule(lr){7-8} \cmidrule(lr){10-11} \cmidrule(lr){13-14} \cmidrule(lr){16-17} \cmidrule(lr){19-20} \cmidrule(l){22-23}  \cmidrule(l){25-26} 
 &  &   &  &   &  & \textit{r}     & RMSE     &  & \textit{r}     & RMSE    &  & \textit{r}      & RMSE      &  & \textit{r}       & RMSE      &  & \textit{r}     & RMSE     &  & \textit{r}     & RMSE   &  & \textit{r}     & RMSE   \\ \midrule
103\_ERICSSON         & SWB   & se   & 54   & 648   &  & 0.85          & 0.38          &  & 0.36           & 0.66    &  & 0.54          & 0.60          &  & 0.28             & 0.68      &  & \textbf{0.87} & \textbf{0.34} &  & 0.78           & 0.45    &  & 0.26            & 0.69     \\
104\_ERICSSON         & NB    & se   & 55   & 660   &  & 0.77          & 0.47          &  & 0.64           & 0.57    &  & 0.68          & 0.55          &  & 0.13             & 0.74      &  & \textbf{0.91} & \textbf{0.31} &  & 0.76           & 0.49    &  & 0.39            & 0.69     \\
203\_FT\_DT           & SWB   & fr   & 54   & 216   &  & \textbf{0.92} & \textbf{0.36} &  & 0.68           & 0.69    &  & 0.47          & 0.82          &  & 0.64             & 0.72      &  & 0.91          & 0.38          &  & 0.79           & 0.57    &  & 0.59            & 0.75     \\
303\_OPTICOM          & SWB   & en   & 54   & 216   &  & 0.92          & 0.33          &  & 0.85           & 0.44    &  & 0.71          & 0.59          &  & 0.43             & 0.76      &  & \textbf{0.93} & \textbf{0.31} &  & 0.71           & 0.59    &  & 0.42            & 0.76     \\
403\_PSYTECHNICS      & SWB   & en   & 48   & 1152  &  & 0.91          & 0.36          &  & 0.81           & 0.50    &  & 0.77          & 0.54          &  & 0.78             & 0.53      &  & \textbf{0.96} & \textbf{0.24} &  & 0.92           & 0.34    &  & 0.73            & 0.57     \\
404\_PSYTECHNICS      & NB    & en   & 48   & 1151  &  & 0.77          & 0.39          &  & 0.82           & 0.35    &  & 0.74          & 0.41          &  & 0.14             & 0.61      &  & \textbf{0.86} & \textbf{0.31} &  & 0.67           & 0.46    &  & 0.55            & 0.51     \\
503\_SWISSQUAL        & SWB   & de   & 54   & 216   &  & 0.92          & 0.34          &  & 0.71           & 0.62    &  & 0.61          & 0.70          &  & 0.59             & 0.71      &  & \textbf{0.94} & \textbf{0.29} &  & 0.85           & 0.46    &  & 0.65            & 0.67     \\
504\_SWISSQUAL        & NB    & de   & 49   & 196   &  & \textbf{0.92} & \textbf{0.37} &  & 0.83           & 0.50    &  & 0.79          & 0.56          &  & 0.54             & 0.77      &  & 0.87          & 0.45          &  & 0.73           & 0.63    &  & 0.60            & 0.73     \\
603\_TNO              & SWB   & nl   & 48   & 192   &  & 0.89          & 0.44          &  & 0.83           & 0.53    &  & 0.69          & 0.69          &  & 0.59             & 0.77      &  & \textbf{0.95} & \textbf{0.29} &  & 0.86           & 0.48    &  & 0.47            & 0.84     \\
ERIC\_FIELD\_GSM\_US  & NB    & en   & 372  & 372   &  & \textbf{0.79} & \textbf{0.36} &  & 0.42           & 0.54    &  & 0.17          & 0.58          &  & 0.60             & 0.47      &  & 0.75          & 0.39          &  & 0.71           & 0.42    &  & 0.51            & 0.51     \\
HUAWEI\_2             & NB    & zh   & 24   & 576   &  & \textbf{0.98} & \textbf{0.21} &  & 0.93           & 0.35    &  & 0.79          & 0.59          &  & 0.63             & 0.75      &  & 0.94          & 0.32          &  & 0.89           & 0.44    &  & 0.97            & 0.24     \\
ITU\_SUPPL23\_EXP1o   & NB    & en   & 44   & 176   &  & 0.92          & 0.31          &  & 0.90           & 0.34    &  & \textbf{0.98} & \textbf{0.15} &  & 0.73             & 0.53      &  & 0.91          & 0.32          &  & 0.91           & 0.33    &  & 0.86            & 0.39     \\
ITU\_SUPPL23\_EXP3d   & NB    & ja   & 50   & 200   &  & 0.92          & 0.27          &  & 0.93           & 0.26    &  & \textbf{0.97} & \textbf{0.17} &  & 0.68             & 0.50      &  & 0.85          & 0.36          &  & 0.84           & 0.36    &  & 0.79            & 0.41     \\
ITU\_SUPPL23\_EXP3o   & NB    & en   & 50   & 200   &  & 0.91          & 0.30          &  & 0.91           & 0.30    &  & \textbf{0.98} & \textbf{0.15} &  & 0.79             & 0.45      &  & 0.88          & 0.35          &  & 0.87           & 0.36    &  & 0.78            & 0.45     \\
TUB\_AUS             & FB    & en   & 50   & 600   &  & \textbf{0.91} & \textbf{0.21} &  & 0.62           & 0.40    &  & 0.65          & 0.39          &  & 0.70             & 0.36      &  & 0.88          & 0.24          &  & 0.73           & 0.35    &  & 0.63            & 0.40      \\
TUB\_LIKE                & SWB   & de   & 8    & 96    &  & 0.98          & 0.25          &  & 0.85           & 0.60    &  & 0.85          & 0.61          &  & 0.59             & 0.93      &  & \textbf{0.99} & \textbf{0.16} &  & 0.89           & 0.53    &  & 0.81            & 0.67     \\
NISQA\_VAL\_LIVE      & FB    & en   & 200  & 200   &  & \textbf{0.82} & \textbf{0.40} &  & 0.42           & 0.64    &  & 0.51          & 0.61          &  & 0.36             & 0.66      &  & 0.67          & 0.52          &  & -0.22          & 0.69    &  & 0.66            & 0.53     \\
NISQA\_VAL\_SIM       & FB    & en   & 2500 & 2500  &  & \textbf{0.90} & \textbf{0.48} &  & 0.45           & 0.99    &  & 0.54          & 0.93          &  & 0.30             & 1.05      &  & 0.86          & 0.56          &  & 0.36           & 1.03    &  & 0.78            & 0.69     \\
NISQA\_TEST\_P501     & FB    & en   & 60   & 240   &  & \textbf{0.95} & 0.31          &  & 0.72           & 0.67    &  & 0.73          & 0.66          &  & 0.80             & 0.59      &  & \textbf{0.95} & \textbf{0.30} &  & 0.80           & 0.59    &  & 0.80            & 0.58     \\
NISQA\_TEST\_NSC      & FB    & de   & 60   & 240   &  & \textbf{0.97} & \textbf{0.23} &  & 0.69           & 0.67    &  & 0.62          & 0.74          &  & 0.78             & 0.59      &  & 0.93          & 0.35          &  & 0.79           & 0.57    &  & 0.78            & 0.59     \\
NISQA\_TEST\_FOR      & FB    & en   & 60   & 240   &  & \textbf{0.95} & \textbf{0.26} &  & 0.52           & 0.71    &  & 0.54          & 0.70          &  & 0.81             & 0.49      &  & 0.92          & 0.33          &  & 0.75           & 0.55    &  & 0.68            & 0.61     \\
NISQA\_TEST\_LIVETALK & FB    & de   & 58   & 232   &  & \textbf{0.90} & \textbf{0.35} &  & 0.70           & 0.58    &  & 0.56          & 0.68          &  & 0.66             & 0.61      &  & N/A           & N/A           &  & N/A            & N/A     &  & N/A             & N/A       \\   \bottomrule
\end{tabular} 

}
\end{table*}

\begin{table*}[!htbp]
\caption{Per-condition validation and test results of the speech quality dimensions in terms of PCC and RMSE after first-order mapping.}
\centering
\label{tab:results_dim}
\resizebox{\textwidth}{!}{%
\begin{tabular}{@{}lcclccccclccccclccccclcc@{}}
\toprule
Dataset               & \multicolumn{5}{c}{NOISINESS}                                 &  & \multicolumn{5}{c}{COLORATION}                                 &  & \multicolumn{5}{c}{DISCONTINUITY}                                 &  & \multicolumn{5}{c}{LOUDNESS}                                \\ \cmidrule(lr){2-6} \cmidrule(lr){8-12} \cmidrule(lr){14-18} \cmidrule(l){20-24} 
                      & \multicolumn{2}{c}{NISQA} &  & \multicolumn{2}{c}{DIAL} &  & \multicolumn{2}{c}{NISQA} &  & \multicolumn{2}{c}{DIAL} &  & \multicolumn{2}{c}{NISQA} &  & \multicolumn{2}{c}{DIAL} &  & \multicolumn{2}{c}{NISQA} &  & \multicolumn{2}{c}{DIAL} \\ \cmidrule(lr){2-3} \cmidrule(lr){5-6} \cmidrule(lr){8-9} \cmidrule(lr){11-12} \cmidrule(lr){14-15} \cmidrule(lr){17-18} \cmidrule(lr){20-21} \cmidrule(l){23-24} 
                      & \textit{r}     & RMSE     &  & \textit{r}     & RMSE    &  & \textit{r}     & RMSE     &  & \textit{r}     & RMSE    &  & \textit{r}     & RMSE     &  & \textit{r}     & RMSE    &  & \textit{r}     & RMSE     &  & \textit{r}     & RMSE    \\ \midrule
503\_SWISSQUAL        & \textbf{0.94} & \textbf{0.26} &  & 0.84           & 0.39    &  & 0.84          & 0.39          &  & \textbf{0.88} & \textbf{0.34} &  & \textbf{0.86} & \textbf{0.31} &  & 0.74           & 0.42    &  & 0.91          & 0.29          &  & \textbf{0.94} & \textbf{0.23} \\
TUB\_AUS             & \textbf{0.97} & \textbf{0.16} &  & 0.88           & 0.29    &  & \textbf{0.84} & \textbf{0.28} &  & 0.81          & 0.3           &  & \textbf{0.92} & \textbf{0.23} &  & 0.61           & 0.46    &  & \textbf{0.74} & \textbf{0.32} &  & 0.62          & 0.38          \\
NISQA\_VAL\_LIVE       & \textbf{0.73} & \textbf{0.49} &  & 0.31           & 0.69    &  & \textbf{0.57} & \textbf{0.43} &  & -0.11         & 0.51          &  & \textbf{0.55} & \textbf{0.56} &  & 0.10           & 0.67    &  & \textbf{0.73} & \textbf{0.47} &  & 0.54          & 0.58          \\
NISQA\_VAL\_SIM       & \textbf{0.86} & \textbf{0.48} &  & 0.40           & 0.87    &  & \textbf{0.84} & \textbf{0.50} &  & 0.25          & 0.90          &  & \textbf{0.84} & \textbf{0.54} &  & 0.23           & 0.97    &  & \textbf{0.81} & \textbf{0.48} &  & 0.38          & 0.76          \\
NISQA\_TEST\_P501     & \textbf{0.95} & \textbf{0.30} &  & 0.86           & 0.48    &  & \textbf{0.91} & \textbf{0.31} &  & 0.62          & 0.60          &  & \textbf{0.91} & \textbf{0.37} &  & 0.68           & 0.65    &  & \textbf{0.95} & \textbf{0.26} &  & 0.88          & 0.40          \\
NISQA\_TEST\_NSC      & \textbf{0.96} & \textbf{0.22} &  & 0.78           & 0.50    &  & \textbf{0.93} & \textbf{0.28} &  & 0.70          & 0.55          &  & \textbf{0.96} & \textbf{0.30} &  & 0.79           & 0.63    &  & \textbf{0.96} & \textbf{0.25} &  & 0.93          & 0.34          \\
NISQA\_TEST\_FOR      & \textbf{0.95} & \textbf{0.23} &  & 0.70           & 0.50    &  & \textbf{0.94} & \textbf{0.24} &  & 0.67          & 0.53          &  & \textbf{0.97} & \textbf{0.25} &  & 0.81           & 0.55    &  & \textbf{0.96} & \textbf{0.20} &  & 0.87          & 0.36          \\
NISQA\_TEST\_LIVETALK & \textbf{0.76} & \textbf{0.47} &  & N/A            & N/A     &  & \textbf{0.87} & \textbf{0.31} &  & N/A           & N/A           &  & \textbf{0.83} & \textbf{0.40} &  & N/A            & N/A     &  & \textbf{0.71} & \textbf{0.36} &  & N/A           & N/A        \\   \bottomrule
\end{tabular}

}
\end{table*}

The results for different pooling mechanism can be seen in Table~\ref{tab:pool}, where a CNN as framewise and SA as time-dependency model is applied. The performance difference between the analysed pooling mechanism is only marginal, however, Attention-Pooling slightly performs better than Average- or Max-Pooling.

\section{Results}
The final model was trained on a set of 59 training and 18 validation datasets with a batch size of 160, learning rate of 0.001, Adam optimiser and bias-aware loss according to \cite{mittag2021qomex}. After each epoch, the model weights were stored, and the results on the training and validation set were calculated as average PCC across all datasets. The training was stopped after the validation PCC did not increase for more than 10 epochs. The model weights with the best performance on the validation set were selected as the final model. This model was then evaluated on the four independent test sets, which were not considered for training, hyper-parameter tuning or model selection.

Table~\ref{tab:results_val_de} presents the validation and test set results of the overall MOS prediction compared to the single-ended models P.563 \cite{p563}, ANIQUE+ \cite{aniquep}, WAWEnets \cite{wenet}, and the double-ended models POLQA \cite{P863}, DIAL \cite{cote}, and VISQOL (v3.1.0) \cite{Chinen2020ViSQOLVA}\footnote{P.563 and ANIQUE+ only allow to predict narrowband signals. Therefore, all signals have been downsampled to 8\,kHz sample rate for the prediction with these models. VISQOL was applied in speech mode (as it gave better results than the audio mode) that only considers frequencies up to 8\,kHz (wideband). Dataset NISQA\_TEST\_LIVETALK contains no reference signals and can therefore only be compared to single-ended models.}. NISQA outperforms the other single-ended speech quality models on most of the datasets, except for the ITU-T Suppl. 23 datasets, where ANIQUE+ achieved the best results. However, it should be noted that these datasets were available for the development of ANIQUE+. POLQA outperforms NISQA on most of the POLQA-Pool datasets (103--603) that contain typical ITU-T P.800 double sentences with a silent pause in between. In contrast, NISQA achieves better results than POLQA on the NISQA test datasets that contain conversational speech.

Table~\ref{tab:results_dim} shows the results of the speech quality dimensions prediction for the datasets for which subjective speech quality dimension ratings are available. NISQA outperforms the double-ended model DIAL on most of the datasets and achieves overall good results with RMSEs of 0.16--0.56.

\section{Conclusions}
We presented the speech quality model NISQA, which, besides overall MOS, also predicts the four speech quality dimensions Noisiness, Coloration, Discontinuity, and Loudness. With this degradation decomposition approach, more insights into the cause of an underlying quality impairment are provided. The model is based on a CNN with following Self-Attention network for time-dependency modelling and an Attention-Pooling block for final time pooling. The model is trained and evaluated on a large set of 81 datasets from different sources and showed to give reliable results on unknown data and real, live phone calls. Furthermore, we open-source the code, the model weights, and speech quality datasets. The presented model is focused on distortions that occur in modern speech communication networks. However, the model weights can also be used to fine-tune the model for related tasks, such as the prediction of enhanced or synthesised speech as shown in \cite{mittag2020b}.

\bibliographystyle{IEEEtran}
\bibliography{mybib}

\end{document}